\documentclass[%
 reprint,
superscriptaddress,
 amsmath,amssymb,
 aps,
]{revtex4-1}
\usepackage{ragged2e} 
\usepackage{booktabs}
\usepackage{multirow}
\usepackage{rotating}
\usepackage{subcaption}

\usepackage{float}
\usepackage{caption}
\usepackage{bm} 
\usepackage{graphicx}
\usepackage{dcolumn}
\usepackage{bm}
\usepackage{hyperref}

\begin{document}

\preprint{APS/123-QED}

\title{Quasinormal Modes and Greybody Factors of Scalar Field Perturbations in the NED Corrected Charged Black Hole Spacetime}
\author{Jie Liang}
\affiliation{%
 College of Physics,Guizhou University,Guiyang,550025,China
}%

\author{Dong Liu}
\affiliation{Department of Physics, Guizhou Minzu University, Guiyang, 550025, China}

\author{Zheng-Wen Long}%
\email{zwlong@gzu.edu.cn (corresponding author)}
\affiliation{%
 College of Physics,Guizhou University,Guiyang,550025,China
}%


\begin{abstract}
Inspired by the quark-antiquark confinement potential, Mazharimousavi et al. \cite{Mazharimousavi:2023okd} proposed a nonlinear electrodynamics (NED) model, and based on this model, they constructed a charged black hole solution that includes a logarithmic correction term ($\propto \frac{\zeta \ln r}{r}$). On the basis of the Reissner-Nordström metric, this solution realizes a long-range confinement correction by introducing the NED parameter $\zeta$, providing a new theoretical perspective for explaining the anomalies in galaxy rotation curves. To deeply explore the dynamic properties of this black hole solution, this paper combines two complementary methods, namely, time-domain evolution and the WKB approximation, to calculate the quasinormal mode (QNM) spectrum of its scalar field perturbations. The research results show that the oscillation frequencies and decay rates of the low-order QNM modes decrease monotonically with the increase of the NED parameter $\zeta$, and exhibit an approximately linear dependence. The analysis of the greybody factor (GF) indicates that as $\zeta$ increases, the transmittance of the low-frequency scalar field also increases. The enhanced long-range confinement effect caused by the increase of $\zeta$ makes low-frequency perturbations more likely to survive and propagate in space-time on the one hand, and at the same time enhances the transmission ability of the low-frequency scalar field. These characteristics provide key theoretical predictions and potential observational features for testing and constraining such NED models in a strong gravitational field environment in the future using the observational data of gravitational wave astronomy or Hawking radiation.
\end{abstract}

\maketitle


\section{Introduction}

In recent years, black hole research has achieved a leap from theory to observation. A series of detections of gravitational wave signals produced by binary black hole mergers by LIGO-Virgo-KAGRA \cite{LIGOScientific:2016aoc, LIGOScientific:2016sjg, LIGOScientific:2017vwq, KAGRA:2023pio} and the successful capture of the shadow image of the M87* black hole by the Event Horizon Telescope \cite{EventHorizonTelescope:2019dse} have provided direct observational evidence for this core prediction of general relativity. At the same time, it provides a unique window for precisely testing GR in the limit of strong gravitational fields and exploring new physics beyond the standard model \cite{LIGOScientific:2021qlt, LIGOScientific:2020stg}.

In the final stage of the Binary Black Hole (BBH) merger, that is, the "ringdown" stage, the merged black hole returns to an equilibrium state by radiating a series of QNMs with specific complex frequencies $\omega = \omega_R + i \omega_I$ ~\cite{Vishveshwara:1970zz, Berti:2009kk, Chandrasekhar:1984siy}. Among them, $\omega_R$ represents the oscillation frequency, and $\omega_I$ represents the damping rate. For the ringdown signals of BBH mergers considering environmental effects \cite{Spieksma:2024voy}, existing and planned gravitational wave detectors can still reliably extract the mass and spin properties of the remnant black hole, being basically unaffected by the surrounding environment. Precise measurement of the QNM spectrum has become an important tool for detecting black hole parameters and testing gravitational theories~\cite{Konoplya:2011qq, Cardoso:2016rao, Liu:2022ygf, Liu:2021xfb, Liang:2024geh, Sirera:2023pbs, Tattersall:2019pvx,Das:2023ess}. In addition, the GFs, which describe the interaction between the perturbation waves and the black hole potential barrier~\cite{Hawking:1975vcx}, have recently been proposed as an independent probe to complement the QNM ringdown test. GFs may be related to the excitation amplitudes of high-frequency QNMs~\cite{Oshita:2022pkc, Oshita:2023cjz, Okabayashi:2024qbz}, the properties of QNMs~\cite{Rosato:2024arw, Oshita:2024fzf}, and the echo signals of horizonless ultracompact objects~\cite{Konoplya:2024lir, Konoplya:2024vuj}, highlighting their potential value in strong gravity tests. 

Although GR is extremely successful at the macroscopic scale and in the weak-field approximation, its prediction of the existence of a spacetime singularity at the center of a black hole \cite{Hawking:1970zqf} suggests that it may be incomplete under extreme conditions. To this end, coupling GR with NED has become an important research direction \cite{Plebanski:1970zz}. NED was originally proposed by Born \cite{Born:1934gh} to solve the self-energy divergence problem of point charges in classical electrodynamics, and its core idea is to modify Maxwell's equations in the strong-field region. When properly coupled with GR, NED can transform a singular black hole into a regular solution with a smooth spacetime center. A milestone achievement in this field is the work of Ayon-Beato and Garcia \cite{Ayon-Beato:2000mjt}, who proved that the historically famous Bardeen regular black hole is exactly the exact solution of the coupling between GR and a specific form of NED. This discovery not only provides a reliable physical basis for the Bardeen model, but also greatly stimulates the enthusiasm for constructing and studying regular black holes using NED \cite{Fan:2016hvf,Rodrigues:2015ayd,Bronnikov:2017sgg,Burinskii:2002pz}. At the same time, exploring the spacetime structure, thermodynamic properties, and observable features of various NED black hole models is crucial for understanding these models beyond the standard Einstein-Maxwell theory, and may provide clues for distinguishing different gravitational or matter field theories \cite{Kumar:2020yem,Atamurotov:2021imh,Sucu:2024xck,Shaymatov:2023jfa,KumarWalia:2024yxn,Rehman:2024jqg,Waseem:2025qoo} 

In this context, this paper focuses on the black hole solution recently obtained by Mazharimousavi under the coupling of NED \cite{Mazharimousavi:2023okd}. Different from many other NED models aimed at solving the singularity problem, the author attempts to phenomenologically simulate the confinement-like interaction potential of the quark-antiquark system in particle physics within the framework of classical field theory, in the form of $U(r) \propto q/r - f \ln(r/r_0)$, which exhibits a Coulomb behavior under the small-distance approximation but shows a logarithmic growth trend at larger distances. Based on this preset form of the electric potential, the specific NED Lagrangian necessary to generate this electric potential is determined inversely. Then, when this specific NED model is coupled with Einstein gravity, the resulting metric function $f(r)$ contains, in addition to the standard Reissner-Nordström (R-N) term, a correction term of the form $\propto \zeta \ln(r)/r$, where $\zeta$ is a coupling parameter directly related to the strength of the logarithmic term of NED. $\ln(r)/r$ is the core feature of this black hole solution, and its physical origin and slow decay property make it different from many other NED black hole models. The literature \cite{Mazharimousavi:2023okd} has pointed out that this logarithmic correction modifies the effective Newtonian gravitational potential acting on a distant test particle, causing its orbital velocity to decrease with the increase of the radius more slowly than that of the R-N black hole. At the same time, we believe that it may exert a unique influence on the dynamic response of spacetime. Therefore, studying the QNMs and greybody factors of this model and systematically examining how the NED parameter $\zeta$ regulates these dynamic properties has become an interesting and important issue. By carefully comparing it with the results of the standard R-N black hole, we expect to reveal the unique dynamic "fingerprint" of this logarithmic correction inspired by the confinement-like potential, which not only helps to deepen the understanding of this specific NED model, but may also provide potential theoretical basis and distinguishable features for searching and verifying such non-standard black hole models in future gravitational wave or electromagnetic wave observations.

The structure of this paper is arranged as follows: Section \ref{sec:metric} introduces the specific form and spacetime metric of the logarithmically corrected NED black hole model. Section \ref{sec:scsle} tests the perturbation of the scalar field in this model. Sections \ref{sec:QNMs} and \ref{sec:numeries} introduce the calculation method and numerical results of QNMs, and analyze the variation law of the frequency with $\zeta$. In Section \ref{sec:GBs}, the greybody factors are analyzed. Section \ref{sec:conclusion} is the summary of the full text. In this paper, we adopt the natural unit system, that is ($c = G = \hbar = 1$). 

\section{Nonlinear Electrodynamics Black Hole Spacetime}
\label{sec:metric}

The NED model proposed by Mazharimousavi et al. \cite{Mazharimousavi:2023okd} has the following Lagrangian:

\begin{equation}
\mathcal{L}=-\frac{16\left( 3\sqrt{-2\mathcal{F}}+\zeta \left( \zeta +\sqrt{\zeta ^{2}+4\sqrt{-2\mathcal{F}}}\right) \right) \sqrt{-2\mathcal{F}}}{3\left( \zeta +\sqrt{\zeta ^{2}+4\sqrt{-2\mathcal{F}}}\right) ^{4}}\mathcal{F}
\end{equation}
where $\mathcal{F}=\frac{1}{4}F_{\mu \nu }F^{\mu \nu}$ is the electromagnetic field invariant, and $\zeta$ is the nonlinear coupling parameter of the NED model. When $\zeta \rightarrow 0$, the Lagrangian reduces to the standard Maxwell theory $\mathcal{L} = -\mathcal{F}$. The electrostatic potential of this NED model in a flat spacetime background is expressed as:

\begin{equation}
U(r) = \frac{q}{r} - \zeta\sqrt{q}\ln\left(\frac{r}{r_0}\right)
\end{equation}
$r_0$ is a constant. The first term is the standard Coulomb potential, and the second term is a logarithmic correction term. This logarithmic term is related to the long - range effect in some physical scenarios. The electric field corresponding to this potential form formed by a point charge is:

\begin{equation}
E(r) = \frac{q}{r^2} + \zeta\frac{\sqrt{q}}{r}
\end{equation}

By minimally coupling the above NED model with Einstein's gravitational theory, the modified Einstein field equations are obtained through the variational principle. Under the assumption of a spherically symmetric metric:

\begin{equation}
ds^2 = -f(r) dt^2 + \frac{dr^2}{f(r)} + r^2 (d\theta^2 + \sin^2\theta d\phi^2)
\end{equation}

Solving the coupled Einstein - NED field equations, the metric function describing a static spherically symmetric black hole is obtained\cite{Mazharimousavi:2023okd}

\begin{figure}[t]
\centering
\includegraphics[width=0.8\columnwidth]{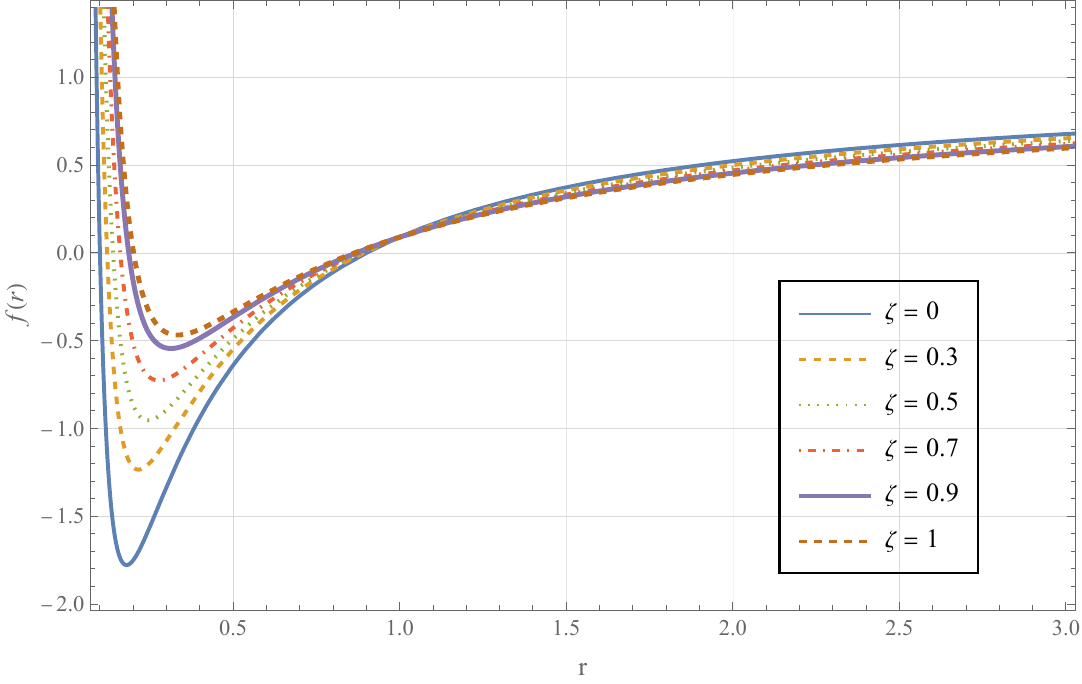}
\caption{\justifying Variation of the metric function $f(r)$ with the radial coordinate $r$ for different values of $\zeta$, where $\zeta = 0$ corresponds to the case of the standard R - N black hole.}
\label{fig:psi(r)}
\end{figure}

\begin{equation}
f(r) = 1 - \frac{2M}{r} + \frac{q^2}{r^2} - \frac{4q\sqrt{q}\zeta}{3r}\ln(r)
\label{eq:metric}
\end{equation}
This metric function describes a black hole spacetime with mass $M$, charge $q$, and modified by the NED parameter $\zeta$. Its properties are jointly determined by these three parameters. When the NED parameter $\zeta = 0$, it degenerates into the standard Reissner - Nordström (R - N) black hole:

\begin{equation}
f(r)=1-\frac{2M}{r}+\frac{q^2}{r^2}
\end{equation}

\begin{figure}[h]
\centering
\includegraphics[width=0.8\columnwidth]{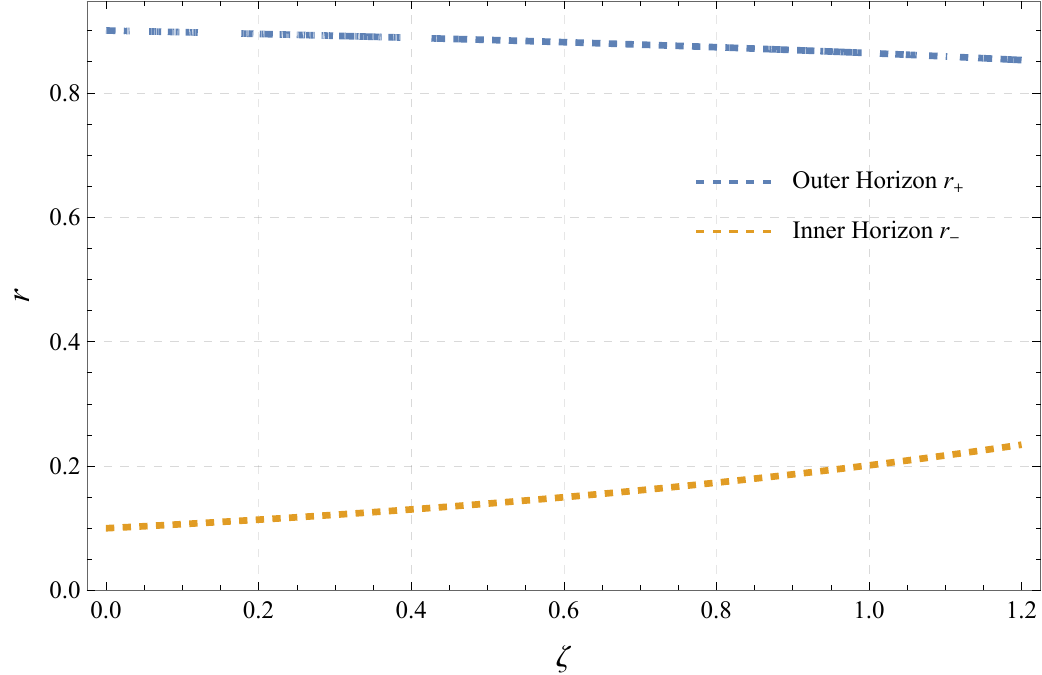}
\caption{\justifying Variation trends of the inner and outer horizons of the black hole with $\zeta$.}
\label{fig:horizon}
\end{figure}

The fourth term of the metric \ref{eq:metric} reflects the nonlinear coupling between the NED correction term and gravity. To explore its specific physical effects, we set the mass - to - charge ratio of the black hole as $M/q = 5/3$ (this ratio ensures the existence of the black hole horizon), and select a set of representative parameters $M = 0.5$ and $q = 0.3$ for specific analysis. In Figure \ref{fig:psi(r)}, we plot the variation trend of the metric function $f(r)$ with the radial coordinate $r$ for different values of $\zeta$. It is observed that the black hole exhibits a double - horizon structure. We set the metric $f(r) = 0$ to determine the horizons of the black hole. The horizon structure of the considered black hole is shown in Figure \ref{fig:horizon}. Compared with the R - N black hole with the same mass and charge, the NED black hole studied in this paper has a larger inner horizon radius $r_-$ and a smaller outer horizon radius $r_+$.

\section{Testing the Perturbation of the Scalar Field}
\label{sec:scsle}

In the background of the metric (\ref{eq:metric}), we discuss the perturbation of a massless test scalar field. The scalar field is denoted as \(\Phi\), and its equation of motion in the framework of general relativity can be written as:

 \begin{equation}
    \frac{1}{\sqrt{-g}} \partial_{\mu} \left( \sqrt{-g} g^{\mu\nu} \partial_{\nu} \Phi \right) = 0,\label{eq:scalar_field}
\end{equation}
The scalar field \(\Phi(t, r, \theta, \phi) = \frac{1}{r} \Psi(t, r) Y_{lm}(\theta, \phi)\) (where \(Y_{lm}(\theta, \phi)\) is the spherical harmonic function, satisfying \(\nabla^2_{\Omega} Y_{lm}(\theta, \phi) = -l(l + 1) Y_{lm}(\theta, \phi)\)). Substituting it into Equation \ref{eq:scalar_field}, the radial equation can be expressed as:

\begin{equation}
-\frac{1}{f(r)} \frac{\partial^2 \Psi}{\partial t^2} + \frac{\partial}{\partial r} \left( f(r) \frac{\partial \Psi}{\partial r} \right) - \frac{l(l + 1) f(r)}{r^2} \Psi = 0.
\label{eq:radial_scalar}
\end{equation}

To simplify the analysis, we introduce the tortoise coordinate \(r_*\):

\begin{equation}
dr_* = \frac{dr}{f(r)},
\label{eq:tortoise}
\end{equation}

The radial equation \eqref{eq:radial_scalar} can be written as a wave equation in the following form:
\begin{equation}
\frac{\partial^2 \Psi}{\partial t^2} - \frac{\partial^2 \Psi}{\partial r_*^2} + V_{\text{eff}}(r) \Psi = 0,
\label{eq:wave}
\end{equation}
where the effective potential can be written as:

\begin{equation}
V_{\text{eff}}(r) = f(r) \left( \frac{l(l+1)}{r^2} + \frac{1}{r} \frac{df(r)}{dr} \right).
\label{eq:potential}
\end{equation}

\begin{figure}[t]
\centering
\includegraphics[width=0.48\columnwidth]{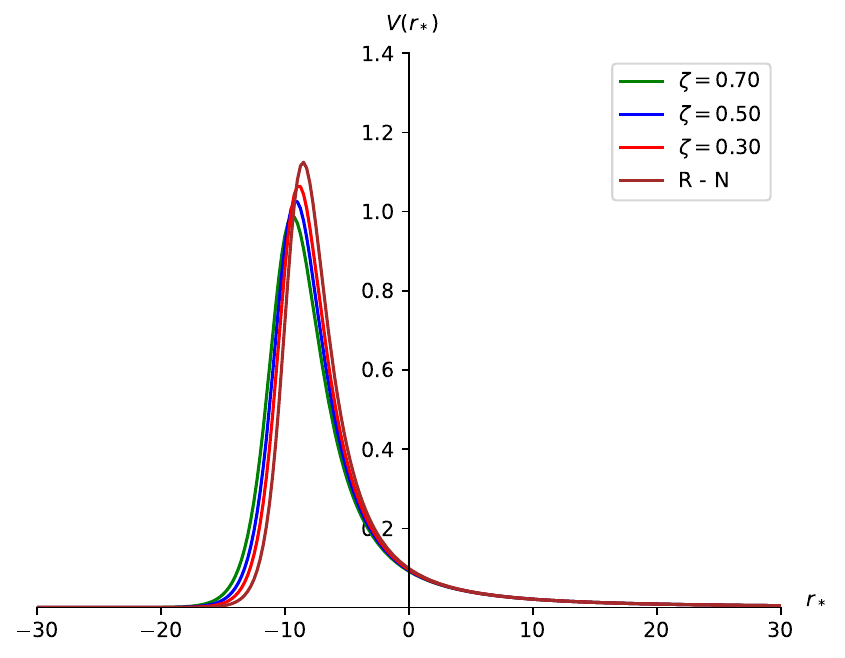}
\hfill 
\includegraphics[width=0.48\columnwidth]{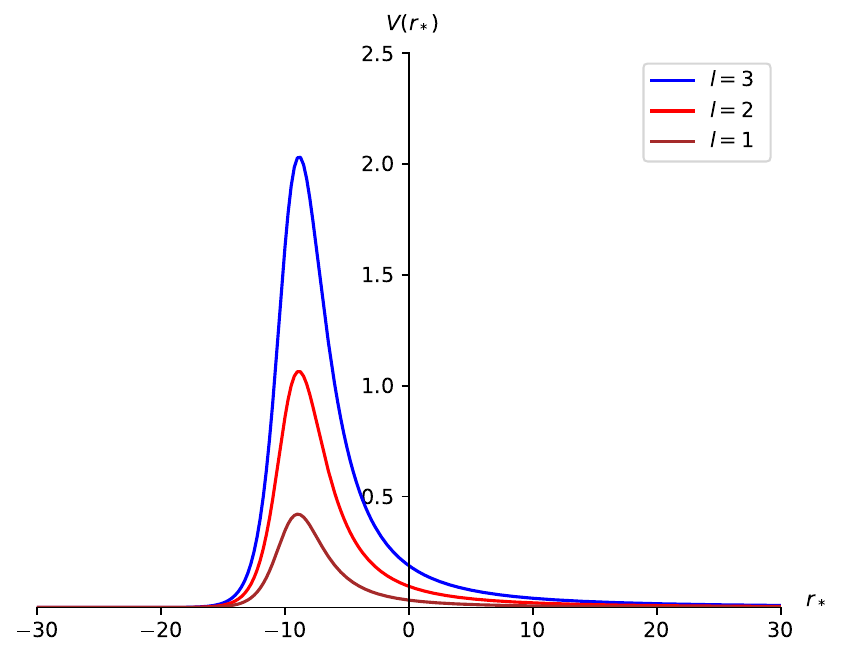} 
\caption{\justifying Variation of the effective potential $V_{\text{eff}}(r)$ of the massless scalar - field perturbation with the tortoise coordinate $r_*$.
Left: Fix the angular quantum number $l = 2$ and show different values of the NED parameter $\zeta$.
Right: Fix the NED parameter $\zeta = 0.3$ and show different values of the angular quantum number $l$.}
\label{fig:Veff_plots} 
\end{figure}

We plot the variation of the effective potential \( V_{\text{eff}} \), as shown in Figure \ref{fig:Veff_plots}. In the left - hand panel, compared with the R - N black hole with \(\zeta = 0\), increasing \(\zeta\) significantly reduces the peak value of the potential barrier and makes the potential barrier wider in the tortoise coordinate. This change means that the incident wave will face a weaker barrier effect and may experience a higher transmittance. We predict that this will lead to a decrease in the real - part frequency \(\omega_R\) of the QNM, because the reduction of the potential - barrier height usually weakens the intensity of the oscillation. At the same time, the broadening of the potential barrier prolongs the residence time of the incident wave in the potential - barrier region, enhancing the "trapping" effect of the perturbation. Therefore, the decay rate \(|\omega_I|\) of the QNM is expected to decrease, and the decay time \(\tau \propto 1/|\omega_I|\) will increase significantly.

In the right - hand panel of Figure \ref{fig:Veff_plots}, we observe that increasing the angular quantum number \( l \) significantly increases the maximum height \( V_{\text{max}} \) of the potential barrier. This result is in line with expectations, because the effective potential \( V_{\text{eff}} \) contains the centrifugal potential - barrier term \( f(r) l(l+1)/r^2 \), and its contribution increases with \( l \). When \( l > 0 \), the centrifugal term becomes the dominant factor determining the height of the potential barrier. We predict that this will lead to an increase in the real - part frequency \(\omega_R\) of the QNM, because a higher potential barrier usually corresponds to a stronger oscillation frequency. In addition, as \( l \) increases, the broadening effect of the potential barrier also becomes more obvious, which may further reduce the decay rate \(|\omega_I|\) and prolong the lifetime of the perturbation. These changes in the effective potential provide an intuitive physical basis for the subsequent study of QNMs.

\section{Calculation of QuasiNormal Modes}
\label{sec:QNMs}

QNMs are the eigenmodes that describe the characteristic oscillations during the relaxation stage of a black hole after it is perturbed. Their complex frequency is $\omega = \omega_R + i \omega_I$. For QNMs, the wave is required to be a pure incoming wave ($\Psi \sim e^{-i \omega r_*}$) at the event horizon when $r \to r_+$ ($r_* \to -\infty$), and a pure outgoing wave ($\Psi \sim e^{+i \omega r_*}$) at spatial infinity when $r \to \infty$ ($r_* \to +\infty$).

In order to accurately calculate the QNM frequencies of the charged black hole under the NED correction, this paper mainly employs two widely used methods: the eigenvalue extraction method based on time-domain evolution and the semi-analytical WKB approximation method. These two methods can verify each other, thus ensuring the reliability of the results.

Firstly, we use the time-domain method to directly perform numerical evolution on the original wave equation \eqref{eq:wave}. To improve numerical stability, we introduce light-cone coordinates $u = t - r_*$ and $v = t + r_*$. Then the wave equation \eqref{eq:wave} is transformed into:

\begin{equation}
    4 \frac{\partial^2 \Psi}{\partial u \partial v} + V_{\text{eff}}(r(u,v)) \Psi(u,v) = 0,
    \label{eq:lightcone_wave}
\end{equation}
The characteristic line integration method developed by Gundlach et al. \cite{Gundlach:1993tp} is used to numerically discretize this equation. Its discretization scheme has a fourth-order high precision $O(\Delta^4)$ on a uniform $(u,v)$ grid:

\begin{equation}
    \Psi_N = \Psi_E + \Psi_W - \Psi_S - \Delta^2 V_{\text{eff}}(r_S) \frac{\Psi_W + \Psi_E}{8} + O(\Delta^4),
    \label{eq:discretization_gundlach}
\end{equation}
where $N$, $W$, $E$, and $S$ represent the values of $\Psi$ at the four vertices $(u + \Delta, v + \Delta)$, $(u + \Delta, v)$, $(u, v + \Delta)$, and $(u, v)$ on the grid respectively, and $\Delta$ is the step size in both the $u$ and $v$ directions. After setting an initial perturbation, time-domain evolution is carried out, and a time-series signal is extracted at a fixed position far from the source. In the later stage of the evolution, this signal is dominated by QNMs. We use the Prony method to fit the signal by expressing it as a sum of a series of decaying exponential functions \cite{Chowdhury:2020rfj,Konoplya:2011qq,Berti:2007dg}:

\begin{equation}
    \Psi(t) \approx \sum_{j=1}^{p} C_j e^{-i \omega_j t},
    \label{eq:prony_fit}
\end{equation}
where $\omega_j$ is the QNM frequency to be extracted. Through least squares fitting, the frequency of the dominant QNM can be directly estimated from the time-domain data.

Then, we use the WKB approximation method for comparison and verification. This method is an elegant and efficient semi-analytical technique, especially with higher accuracy when the angular quantum number $l$ is large \cite{Schutz:1985km,Iyer:1986np,Iyer:1986nq,Konoplya:2003ii}. In this paper, the sixth-order WKB approximation formula is adopted:
\begin{equation}
    \frac{i(\omega^2 - V_{\text{max}})}{\sqrt{-2 V''_{\text{max}}}} - \sum_{k=2}^{6} \Lambda_k(\omega) = n + \frac{1}{2}, \quad (n = 0,1,2,\dots),
    \label{eq:WKB_6th}
\end{equation}
where $n$ is the overtone number of the QNM, $V_{\text{max}} \equiv V_{\text{eff}}(r_{0,\text{peak}})$ is the value of the effective potential at its peak, $V''_{\text{max}} \equiv \frac{d^2 V_{\text{eff}}}{dr_0^2} \Big|_{r_{0,\text{peak}}}$ is the second derivative of the effective potential with respect to the coordinate $r_0$ at the peak, and $\Lambda_k$ represents the $k$-th order correction term \cite{Konoplya:2003ii}. 

\begin{table}[htbp]
\centering
\footnotesize 
\setlength{\heavyrulewidth}{0.12ex} 
\setlength{\lightrulewidth}{0.08ex} 
\setlength{\tabcolsep}{3pt}
\resizebox{\columnwidth}{!}{
\begin{tabular}{ccccc}
\toprule 
$\zeta$ & $\omega_{\text{WKB}}$ & $\omega_{\text{Prony}}$ & $\Delta_{\omega_R} (\%)\!$ & $\Delta_{|\omega_I|} (\%)\!$ \\
\midrule %
0.0  & 1.03477 - 0.196664I & 1.03254 - 0.208841I & 0.216 & 5.834 \\
0.1  & 1.02641 - 0.192439I & 1.02648 - 0.202128I & 0.007 & 4.797 \\
0.2  & 1.01797 - 0.188206I & 1.01816 - 0.197526I & 0.019 & 4.719 \\
0.3  & 1.00948 - 0.183965I & 1.00959 - 0.192998I & 0.011 & 4.680 \\
0.4  & 1.00091 - 0.179715I & 1.00109 - 0.188356I & 0.018 & 4.587 \\
0.5  & 0.992268 - 0.175454I & 0.992644 - 0.184021I & 0.038 & 4.656 \\
0.6  & 0.983555 - 0.171181I & 0.983865 - 0.179046I & 0.032 & 4.393 \\
0.7  & 0.974766 - 0.166895I & 0.975142 - 0.174384I & 0.039 & 4.296 \\
0.8  & 0.965899 - 0.162595I & 0.966337 - 0.169722I & 0.045 & 4.197 \\
0.9  & 0.956951 - 0.158279I & 0.95699 - 0.165647I & 0.004 & 4.448 \\
1.0  & 0.947919 - 0.153945I & 0.947726 - 0.160992I & 0.020 & 4.375 \\
\bottomrule 
\end{tabular}
}
\caption{\justifying QNM frequencies (fixed $l=2, n=0$) calculated by WKB and Prony methods and their relative errors} 
\label{tab:WKB_Prony_abs}
\end{table}

\begin{table*}[htbp]
\centering
\small
\setlength{\heavyrulewidth}{0.12ex} 
\setlength{\lightrulewidth}{0.08ex} 
\setlength{\tabcolsep}{4pt} 
\resizebox{0.9\textwidth}{!}{
\begin{tabular}{ccccccccccccc}
\toprule
$(l, n)$ & $\zeta = 0.0$ & $\zeta = 0.1$ & $\zeta = 0.2$ & $\zeta = 0.3$ & $\zeta = 0.4$ & $\zeta = 0.5$ & $\zeta = 0.6$ & $\zeta = 0.7$ & $\zeta = 0.8$ & $\zeta = 0.9$ & $\zeta = 1.0$ \\
\midrule
$(0, 0)$ & 0.23732 & 0.234459 & 0.231691 & 0.229539 & 0.226493 & 0.224108 & 0.220551 & 0.218415 & 0.21555 & 0.212486 & 0.210681 \\
$(1, 0)$ & 0.627041 & 0.621819 & 0.616553 & 0.611248 & 0.605899 & 0.600506 & 0.595067 & 0.589581 & 0.584042 & 0.578451 & 0.572806 \\
$(2, 0)$ & 1.03477 & 1.02641 & 1.01797 & 1.00948 & 1.00091 & 0.992268 & 0.983555 & 0.974766 & 0.965899 & 0.956951 & 0.947919 \\
$(3, 0)$ & 1.44477 & 1.43318 & 1.42151 & 1.40974 & 1.39788 & 1.38592 & 1.37386 & 1.36169 & 1.34941 & 1.33703 & 1.32453 \\
\midrule
$(0, 1)$ & 0.196903 & 0.194652 & 0.192711 & 0.192178 & 0.189722 & 0.188831 & 0.185273 & 0.184964 & 0.182847 & 0.179885 & 0.179733 \\
$(1, 1)$ & 0.575161 & 0.572111 & 0.568974 & 0.565788 & 0.562523 & 0.559182 & 0.555768 & 0.552264 & 0.548648 & 0.544932 & 0.541105 \\
$(2, 1)$ & 0.998742 & 0.991961 & 0.985093 & 0.978134 & 0.97108 & 0.963927 & 0.95667 & 0.949303 & 0.94182 & 0.934214 & 0.926478 \\
\bottomrule
\end{tabular}
}
\caption{Variation of the real part of QNM frequency $\omega_R$ with parameter $\zeta$}
\label{tab:qnm_real_part}
\end{table*}
\begin{table*}[htbp]
\centering
\small
\setlength{\heavyrulewidth}{0.12ex} 
\setlength{\lightrulewidth}{0.08ex} 
\setlength{\tabcolsep}{4pt} 
\resizebox{0.9\textwidth}{!}{
\begin{tabular}{ccccccccccccc}
\toprule
$(l, n)$ & $\zeta = 0.0$ & $\zeta = 0.1$ & $\zeta = 0.2$ & $\zeta = 0.3$ & $\zeta = 0.4$ & $\zeta = 0.5$ & $\zeta = 0.6$ & $\zeta = 0.7$ & $\zeta = 0.8$ & $\zeta = 0.9$ & $\zeta = 1.0$ \\
\midrule
$(0, 0)$ & 0.20451 & 0.200119 & 0.195652 & 0.190681 & 0.18645 & 0.181675 & 0.177837 & 0.172822 & 0.168346 & 0.16392 & 0.158396 \\
$(1, 0)$ & 0.198442 & 0.194104 & 0.189763 & 0.185417 & 0.181066 & 0.176707 & 0.172341 & 0.167965 & 0.163578 & 0.159179 & 0.154764 \\
$(2, 0)$ & 0.196664 & 0.192439 & 0.188206 & 0.183965 & 0.179715 & 0.175454 & 0.171181 & 0.166895 & 0.162595 & 0.158279 & 0.153945 \\
$(3, 0)$ & 0.196178 & 0.191983 & 0.18778 & 0.183568 & 0.179344 & 0.17511 & 0.170863 & 0.166602 & 0.162326 & 0.158032 & 0.15372 \\
\midrule
$(0, 1)$ & 0.688695 & 0.67366 & 0.657675 & 0.637193 & 0.623158 & 0.604084 & 0.593529 & 0.572334 & 0.556455 & 0.542063 & 0.518007 \\
$(1, 1)$ & 0.619057 & 0.604811 & 0.590609 & 0.576404 & 0.562217 & 0.548037 & 0.533849 & 0.519659 & 0.505472 & 0.49126 & 0.477016 \\
$(2, 1)$ & 0.599584 & 0.586405 & 0.573215 & 0.560012 & 0.546793 & 0.533554 & 0.520291 & 0.507002 & 0.49368 & 0.480321 & 0.466917 \\
\bottomrule
\end{tabular}
}
\caption{Variation of the imaginary part of QNM frequency $|\omega_I|$ with parameter $\zeta$ ($\omega_I < 0$)}
\label{tab:qnm_imag_part}
\end{table*}

\section{Numerical Results and Analysis of QuasiNormal Modes for NED Charged Black Holes}
\label{sec:numeries}

\begin{figure}[htbp]
\centering
\includegraphics[width=0.4\columnwidth]{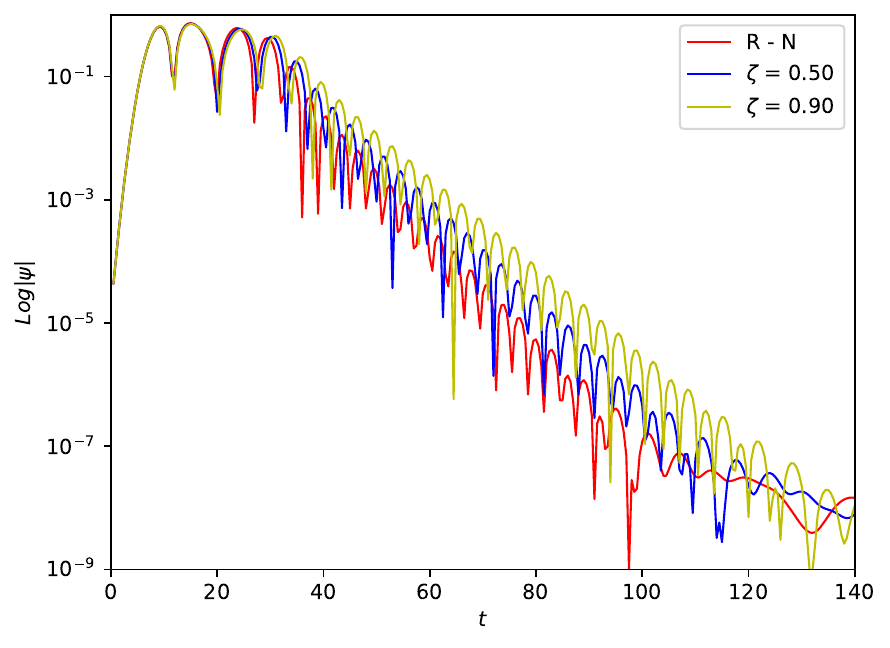}
\includegraphics[width=0.4\columnwidth]{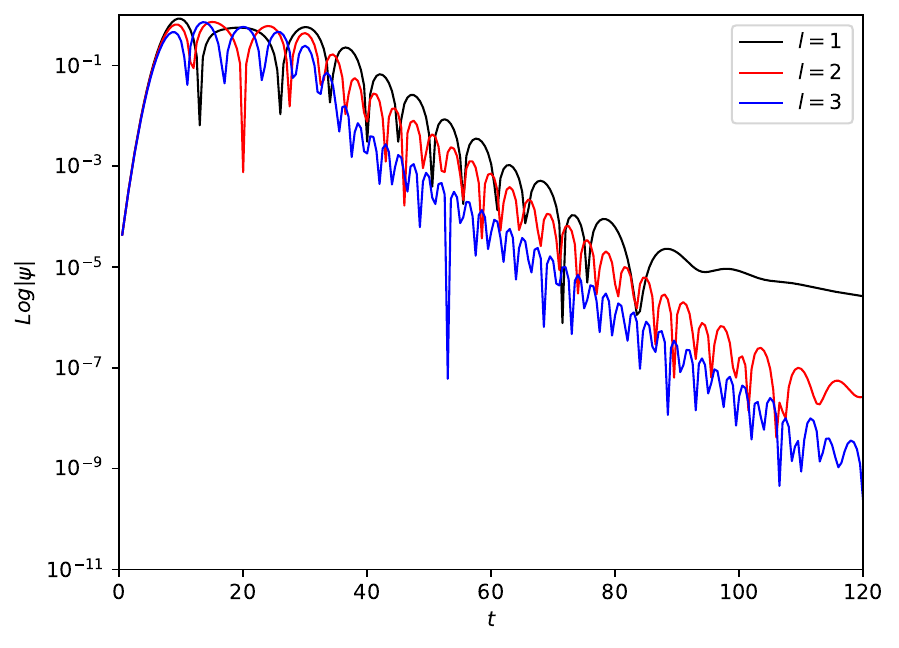}
\caption{\justifying Time-domain evolution of the scalar field wavefunction \(|\Psi(t)|\): The left panel shows the QNM characteristics for different values of \(\zeta\) when \(l=2\), \(M=0.5\), and \(q=0.3\) are fixed; the right panel shows the QNM characteristics for different angular quantum numbers \(l\) (1, 2, 3) when \(\zeta=0.3\), \(M=0.5\), and \(q=0.3\) are fixed.}
\label{fig:waveform}
\end{figure}

To intuitively understand the perturbation dynamics, we first analyze the time-domain evolution of the scalar field through numerical simulations, as shown in Fig.~\ref{fig:waveform}. The scalar field evolution exhibits the classical three-stage characteristics in black hole perturbation theory: an initial transient response (\(t \lesssim 50\)) dominated by the initial conditions, followed by an exponentially decaying phase driven by QNMs, and finally transitioning to a power-law decaying tail at late times. As seen in Fig.~\ref{fig:waveform} (left panel), as the NED parameter \(\zeta\) increases from 0 (R-N limit) to 0.9, the waveform oscillation frequency gradually slows down, and the decay time significantly lengthens.

These features closely correspond to the effective potential \(V_{\text{eff}}\) analyzed in Section~\ref{sec:scsle} (Fig.~\ref{fig:Veff_plots}). Increasing \(\zeta\) lowers the potential barrier peak and broadens it, resulting in a decrease in the QNM frequency while extending the lifetime of the QNM mode. Conversely, increasing \(l\) significantly enhances the centrifugal potential term (\(\propto l(l+1)/r^2\)), raising the barrier height. This correlation between the time-domain evolution and the effective potential provides a physical basis and preliminary validation for accurately extracting QNM frequencies in the following analysis.

Next, to verify the reliability of the calculation method and assess its accuracy, we extracted the QNM frequency of the fundamental mode (\(n=0, l=2\)) from the time-domain data using the Prony method and compared it with the results of the sixth-order WKB approximation, as detailed in Table~\ref{tab:WKB_Prony_abs}. The relative error between the two methods is defined as:
\begin{equation}
    \Delta = \left| \frac{\omega_{\text{Prony}} - \omega_{\text{WKB}}}{\omega_{\text{Prony}}} \right| \times 100\%.
    \label{eq:delta}
\end{equation}

The comparison results show that the oscillation frequencies \(\omega_R\) calculated by the Prony and WKB methods are in high agreement, with the relative error \(\Delta_{\omega_R}\) ranging from only 0.004\% to 0.216\%, with an average of about 0.041\%. However, the calculation of the damping rate \(|\omega_I|\) shows a larger discrepancy, with \(\Delta_{|\omega_I|}\) ranging from 4.2\% to 5.8\%, with an average of about 4.6\%, and reaching its maximum value at the R-N limit (\(\zeta=0\)). This magnitude of error is not unexpected in QNM studies, especially based on the fundamental mode and small \(l\) values. The accuracy of the WKB method for the imaginary part is limited by the local Taylor approximation of the effective potential, while the time-domain method directly fits the decaying tail through global waveform matching, thereby avoiding the inherent error of asymptotic series truncation and enabling more accurate extraction of \(\omega_I\)~\cite{Konoplya:2011qq}.

\begin{figure*}[ht] 
    \centering 
    \begin{subfigure}[b]{0.49\textwidth}
        \centering
       \includegraphics[width=\textwidth]{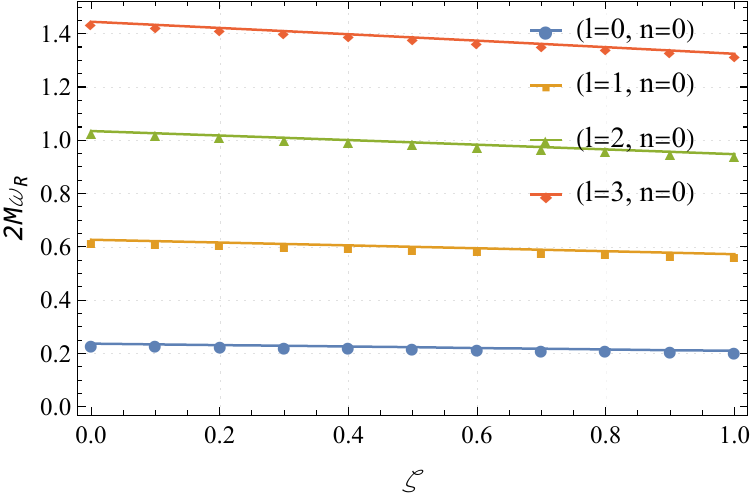}
        \caption{Variation of $2M\omega_R$ with $\zeta$ ($n=0$)} 
        \label{fig:n0_Re_cn} 
    \end{subfigure}
    \hfill 
    \begin{subfigure}[b]{0.49\textwidth} 
        \centering
	   \includegraphics[width=\textwidth]{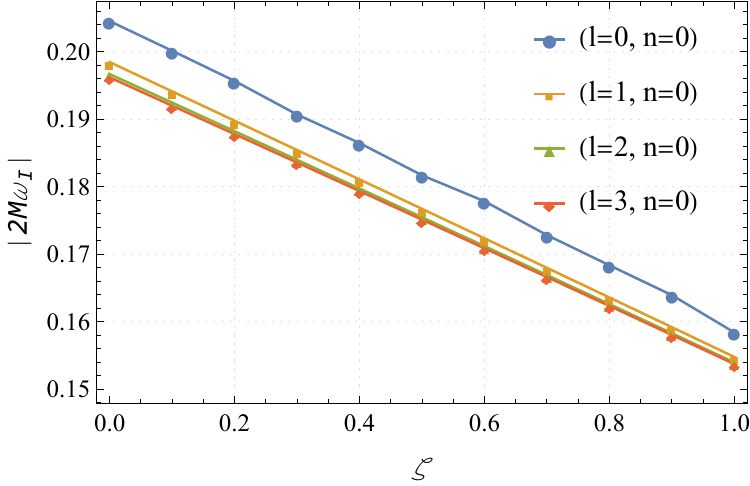}
        \caption{Variation of $2M|\omega_I|$ with $\zeta$ ($n=0$)}
        \label{fig:n0_Im_cn}
    \end{subfigure}
    \caption{\justifying Dependence of the fundamental mode QNM frequency on the NED parameter $\zeta$, showing the cases for different angular quantum numbers $l=0, 1, 2, 3$. Figure (a) shows the normalized oscillation frequency $2M\omega_R$, and Figure (b) shows the absolute value of the normalized damping rate $2M|\omega_I|$. The points at $\zeta=0$ correspond to the values for the Reissner-Nordström black hole.}
    \label{fig:n0_modes_cn} 
\end{figure*}
\begin{figure*}[ht]
    \centering
    \begin{subfigure}[b]{0.48\textwidth}
        \centering
         \includegraphics[width=\textwidth]{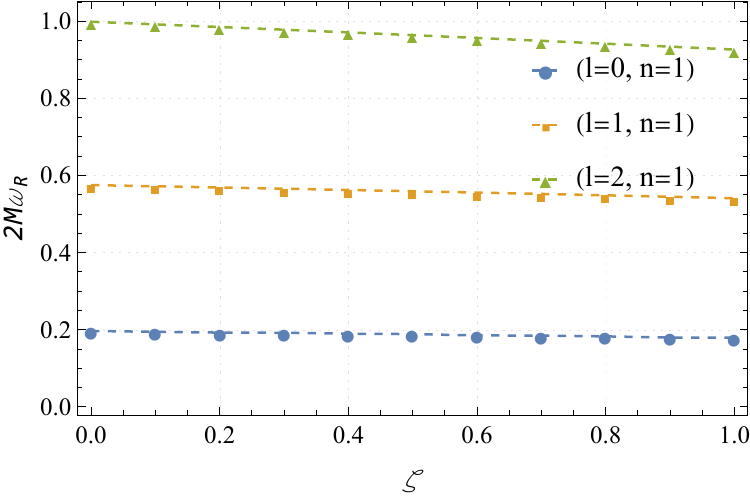}
        \caption{Variation of $2M\omega_R$ with $\zeta$ ($n=1$)}
        \label{fig:n1_Re_cn}
    \end{subfigure}
    \hfill
    \begin{subfigure}[b]{0.48\textwidth}
        \centering
               \includegraphics[width=\textwidth]{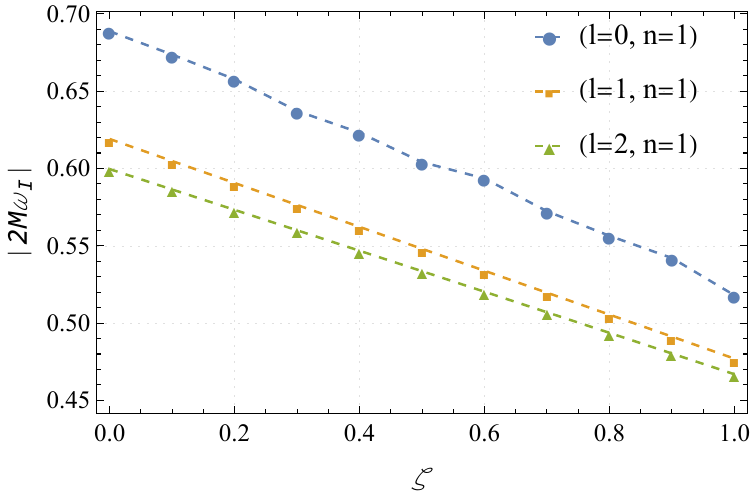}
        \caption{Variation of $2M|\omega_I|$ with $\zeta$ ($n=1$)}
        \label{fig:n1_Im_cn}
    \end{subfigure}
    \caption{Dependence of the first overtone ($n=1$) QNM frequency on the NLED parameter $\zeta$, showing the cases for $l=0, 1, 2$. Figure (a) shows $2M\omega_R$, and Figure (b) shows $2M|\omega_I|$.}
    \label{fig:n1_modes_cn}
\end{figure*}

We calculated the QNM frequencies of the scalar field in the background of the NED-corrected charged black hole using the sixth-order WKB approximation method. The nonlinear parameter $\zeta$ ranges from 0 to 1. The dependence of the real and imaginary parts of the QNM frequencies on different $(l,n)$ is shown in Fig.~\ref{fig:n0_modes_cn} and~\ref{fig:n1_modes_cn}. For both the fundamental mode ($n=0$) and the first overtone ($n=1$), the oscillation frequency $2M\omega_R$ and the damping rate $2M|\omega_I|$ decrease monotonically with increasing $\zeta$ for different angular quantum numbers $l$. This clearly indicates that the NED effect, parameterized by $\zeta$, generally leads to slower oscillation and longer lifetimes of the perturbation modes, which is entirely consistent with the aforementioned analysis based on the change of the effective potential barrier (increasing $\zeta$ lowers and broadens the barrier). At the same time, we also observed that increasing the angular quantum number $l$ increases the oscillation frequency $2M\omega_R$ and decreases the damping rate $2M|\omega_I|$, i.e., the decay becomes slower. Notably, as the value of $l$ increases, the trend of QNM frequency decreasing with $\zeta$ seems to become linear. In the small $\zeta$ limit, we may be able to approximate the behavior of the QNM frequencies through linear fitting or perturbation methods.

Next, we compare the characteristics of the fundamental mode ($n=0$) and the first overtone ($n=1$) (shown in Fig.~\ref{fig:n0_modes_cn} and~\ref{fig:n1_modes_cn} respectively). Consistent with the general physical picture of QNMs, for a fixed angular quantum number $l$, the first overtone ($n=1$) exhibits a significantly larger damping rate $2M|\omega_I|$ than the fundamental mode ($n=0$), i.e., its lifetime is shorter. This reflects that the overtone modes, as higher-order excited states, generally decay more rapidly. As for the oscillation frequency $2M\omega_R$, our calculations show (comparing Fig.~\ref{fig:n0_Re_cn} with Fig.~\ref{fig:n1_Re_cn}) that for a given $l$, the frequency of the fundamental mode is usually slightly higher than or very close to the frequency of the first overtone.

In summary, the QNM frequency results calculated by the WKB method are consistent with the conclusions of the time-domain evolution and effective potential analysis. This indicates that the NED parameter $\zeta$, by changing the structure of the effective potential barrier, systematically reduces the oscillation frequency and decay rate of the scalar field perturbations, thereby extending the mode lifetime. This physical effect is clearly reflected in both the fundamental mode and overtones, and is superimposed on the basic QNM spectral structure determined by the angular quantum number $l$ and the overtone number $n$.

\section{Greybody Factors}
\label{sec:GBs}

\begin{figure}[h]
\centering
\includegraphics[width=0.8\columnwidth]{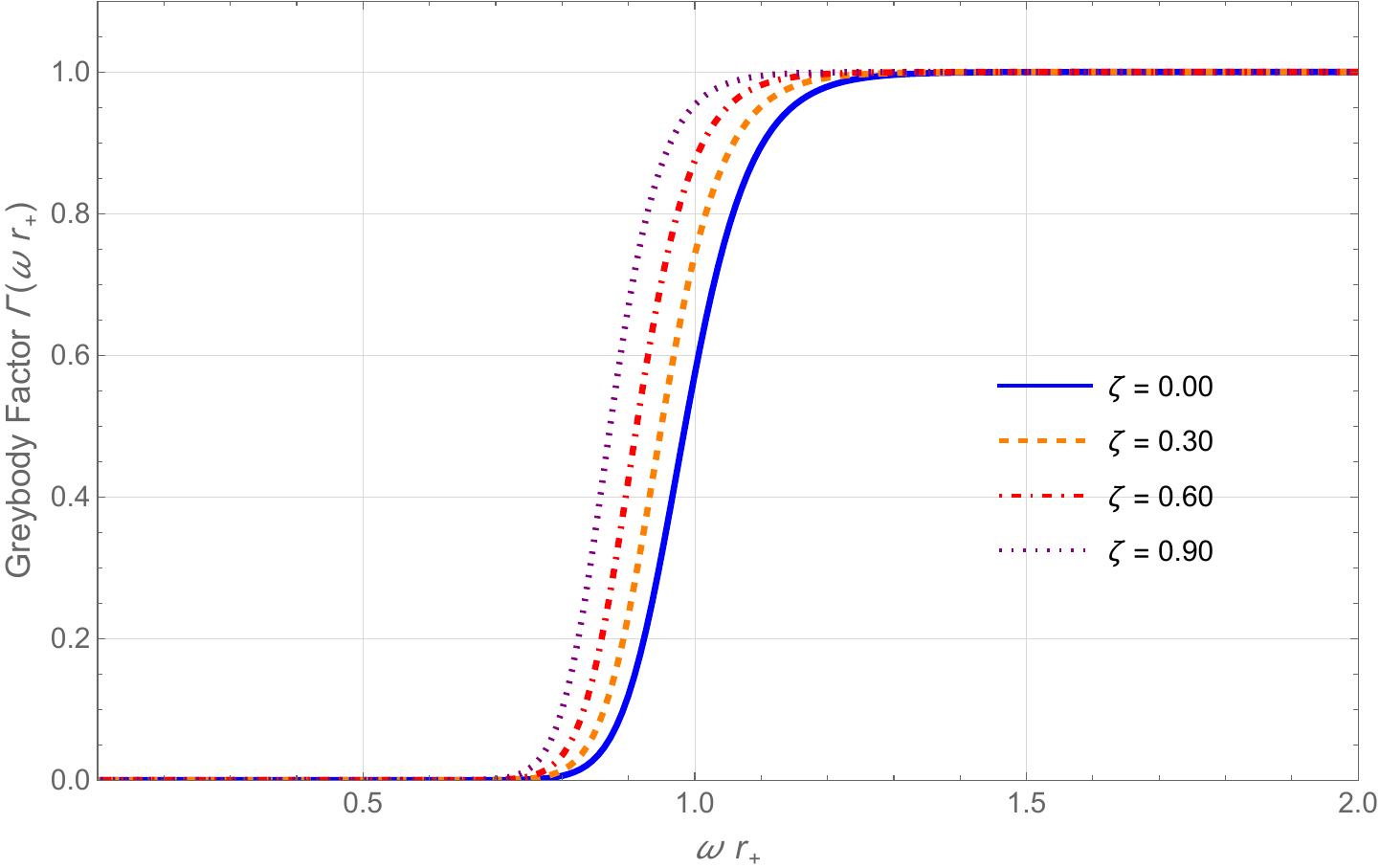}
\caption{\justifying Variation of the greybody factor \(\Gamma(\omega) = |T(\omega)|^2\) for scalar perturbations (\(\ell=2\)) with the normalized frequency \(\omega r_+\). Different curves correspond to different values of the NED parameter \(\zeta\).}
\label{fig:greybody}
\end{figure}

The external curvature and gravitational potential barrier surrounding a black hole affect the spectrum of radiation it emits \cite{Visser:1998ke, Boonserm:2008zg, Sakalli:2016fif}. To quantify this effect, we introduce the frequency-dependent "greybody factor" \(\Gamma(\omega)\), which represents the probability that a quantum with frequency \(\omega\), produced near the event horizon, successfully traverses the potential barrier and propagates to infinity. We employ a semi-analytical WKB approximation method, with the boundary conditions for GF from the literature \cite{Boonserm:2019mon,Kokkotas:1999bd,Boonserm:2008zg}:
\begin{equation}
\Gamma(\omega) \geq \operatorname{sech}^2 \left( \frac{1}{2\omega} \int_{r_h}^\infty V_{\text{eff}}(r) dr \right),
\label{eq:greybody_bound_revised} 
\end{equation}

where \(r_h\) is the event horizon radius. For ease of analysis, we normalize the frequency \(\omega\) by the outer event horizon radius \(r_+\) and plot the greybody factor as a function of \(\omega r_+\) (as shown in Fig.~\ref{fig:greybody}). The angular quantum number is fixed at \(\ell=2\). A significant trend is that the greybody factor increases with the normalized frequency and with \(\zeta\), especially noticeably at (\(\omega r_+ \approx 0.9\)). This enhancement with increasing $\zeta$ is consistent with the change in the effective potential barrier peak \(V_{\text{max}}\) (see Fig.~\ref{fig:Veff_plots}). The lower barrier makes it easier for low-energy scalar waves to penetrate, thereby increasing the transmission coefficient \(\Gamma(\omega)\). On the other hand, when the frequency is sufficiently high (\(\omega r_+ \gtrsim 1.2\)), the greybody factors for all different \(\zeta\) values quickly approach 1. This is because high-energy waves can easily overcome the barrier, and their transmission behavior is then mainly determined by geometric optics, becoming insensitive to the specific structure of the barrier (including the modification introduced by \(\zeta\)).

The analysis of the greybody factor reveals how the NED parameter affects the scattering characteristics of the black hole. Furthermore, it corroborates the deformation of the effective potential barrier and the changing trend of the QNM frequency. The enhancement of the low-frequency greybody factor caused by the NED parameter (\(\zeta > 0\)) suggests that, compared to R-N black holes with the same mass and charge, these modified black holes will more effectively emit low-energy scalar particles through the Hawking radiation process. This predicts that the peak of the observed Hawking radiation spectrum will shift towards the low-frequency end, and the total radiation power may also increase. This unique spectral feature provides a potential window for future astronomical observations to test or constrain such nonlinear electrodynamics models.

\section{Conclusion}
\label{sec:conclusion}

In this paper, we investigated a NED charged black hole model that differs from the standard Maxwell theory, where the correction term in the metric takes a logarithmic form \(\propto \zeta \ln r / r\). We focused on how the NED parameter \(\zeta\) affects the dynamic characteristics of the black hole, as well as its effect on the scattering and radiation properties.

Compared to the standard R-N black hole, the correction parameter $\zeta$ in this model leads to an increase in the inner horizon radius and a decrease in the outer horizon radius. As $\zeta$ increases, the effective potential barrier $V_{\mathrm{eff}}$ experienced by the scalar field decreases monotonically, while the barrier width gradually broadens in the tortoise coordinate space ($r_*$). This barrier deformation directly affects the characteristics of the QNMs: both the oscillation frequency $\omega_R$ and the absolute value of the damping rate $|\omega_I|$ decrease with increasing $\zeta$, leading to a significant increase in the mode lifetime. Within the range of $\zeta$ we studied, the imaginary part of the QNM remains negative, indicating that the black hole model remains stable. The results calculated using the 6th-order WKB method and the time-domain method show that the difference rate between the real parts is about 0.041\%, and the difference rate between the imaginary parts is about 4.6\%, verifying the reliability of the calculations. Subsequently, we plotted the relationship between the QNM frequency and $\zeta$, and found that it exhibits an approximately linear dependence. Furthermore, as the angular quantum number $l$ increases, the linear characteristic becomes more pronounced. This result not only provides a concise analytical framework for the influence of the parameterized model, but also lays the foundation for future constraints on $\zeta$ through gravitational wave observations. At the same time, it supports treating the $\zeta \ln r/r$ term as a perturbation to the R-N spacetime. The dependence of the QNM frequency on the angular quantum number $l$ and $n=0,1$ is consistent with expectations. Consistent with the analysis of the reduced effective potential barrier, the greybody factor $\Gamma(\omega)$ in the low-frequency region systematically increases with increasing $\zeta$.

In summary, this study reveals the significant impact of the logarithmic NED correction on the dynamic characteristics of charged black holes: increasing \(\zeta\) leads to an approximately linear decrease in the QNM frequencies \(\omega_R\) and \(|\omega_I|\), an extended mode lifetime, and an enhanced low-frequency greybody factor \(\Gamma(\omega)\). These features distinguish it from the standard R-N black hole, providing testable signals for future gravitational wave and radiation observations.

\section*{Acknowledgments}
This research partly supported by the
National Natural Science Foundation of China (Grant No. 12265007).

\newpage


\bibliographystyle{unsrt}  
\bibliography{ref}
\bibliographystyle{apsrev4-1}

\end{document}